\documentclass[a4paper]{VisionStyle}
\usepackage{epsfig}
\def\simless{\mathbin{\lower 3pt\hbox
     {$\rlap{\raise 5pt\hbox{$\char'074$}}\mathchar"7218$}}}   
\def\simmore{\mathbin{\lower 3pt\hbox
     {$\rlap{\raise 5pt\hbox{$\char'076$}}\mathchar"7218$}}}   
\def\msun{{\rm M}_{\odot}}                                     

\begin{document}

\title{The XEUS Mission}

\author{Johan Bleeker\inst{1} \and  Mariano M\'{e}ndez\inst{1}} 

\institute{
SRON, National Institute for Space Research, Sorbonnelaan 2, 3584 CA
Utrecht, The Netherlands
}

\maketitle 

\begin{abstract}

{\em XEUS}, the {\em X-ray Evolving Universe Spectroscopy} mission,
constitutes at present an ESA-ISAS initiative for the study of the
evolution of the hot Universe in the post-Chandra/XMM-Newton era. The
key science objectives of XEUS can be formulated as the:

\noindent -- Search for the origin, and subsequent study of growth, of
the first massive black holes in the early Universe.

\noindent -- Assessment of the formation of the first gravitationally
bound dark matter dominated systems, i.e. small groups of galaxies, and
their evolution.

\noindent -- Study of the evolution of metal synthesis up till the
present epoch. Characterization of the true intergalactic medium.

To reach these ambitious science goals the two salient characteristics
of the XEUS observatory entail: 

1. Its effective spectroscopic grasp, combining a sensitive area $> 20$
m$^{2}$ below a photon energy of 2 keV with a spectral resolution
better than 2 eV. This allows significant detection of the most
prominent X-ray emission lines (e.g. O-VII, Si-XIII and Fe-XXV) in
cosmologically distant sources against the sky background.

2. Its angular resolving power, between 2 and 5 arc seconds, to
minimize source confusion as well as noise due to the galactic X-ray
foreground emission.

To accommodate these instrument requirements a mission concept has been
developed featuring an X-ray telescope of 50 meter focal length,
comprising two laser-locked spacecraft, i.e. separate mirror and
detector spacecraft's. The telescope is injected in a low earth orbit
with an inclination commensurate with the ISS, a so-called fellow
traveler orbit. At present an on-orbit growth of the mirror spacecraft
is foreseen through a robotic upgrade with the aid of the ISS, raising
the mirror diameter from 4.5 to 10 meter. The detector spacecraft,
formation flying in a non-Keplerian orbit in tandem with the mirror
spacecraft will be replaced at 5 year intervals after run-out of
consumables with an associated upgrade of the focal plane package.


\end{abstract}

\section{Introduction}

At the end of the 20th century, the promise of high spatial and
spectral resolution in X-rays has become a reality. The two major X-ray
observatories nowadays operational, NASA's Chandra and ESA's XMM
Newton, are providing a new, clear-focused, vision of the X-ray
Universe, in a way that had not been possible in the first 40 years of
X-ray astronomy. These two missions complement each other very well:
Chandra has a $\simless 0.5$ arcsec angular resolution and the
capability of high spectral resolution on a variety of point sources
with the High- and Low-Energy Transmission Grating Spectrometers, while
XMM-Newton has a larger spectroscopic area and bandwidth (up to
$\sim$15 keV), and the capability of high-resolution spectroscopic
observations on spatially extended sources.

Despite their superb capabilities, these two missions cannot be used
for detailed studies of objects at very high redshifts ($z \simmore
5$). At present, the Chandra and XMM-Newton deep surveys in very narrow
pieces of the sky allow us to detect quasars up to redshifts of 6.28,
(\cite{jbleeker-D:bra02}), but we are not able to produce X-ray spectra
of these objects at such distance. 

Both Chandra and XMM-Newton are expected to be operational for the next
ten years. This is the typical timescale for a new mission to be
planned and developed, therefore this is the time to assess what is the
future of X-ray astronomy, and to start planning for Chandra's and
XMM-Newton's follow-up. ESA's response to this challenge has been
cosmology, and the unique role that X-ray astronomy can play in 
studying the formation and evolution of the hot Universe.

\section{Science case}

Some of the basic cosmological issues that need to be resolved from the
observational point of view are:

\begin{itemize}

\item What was the physics of the early Universe?

\item What is the nature of dark matter?

\item How did the large-scale structure form?

\item What did form first, massive black holes or stars and galaxies? 

\item How did galaxies form and evolve, and what was the role of
massive black holes in the centers of galaxies?

\item What is the history of the baryons in the Universe?

\item How and when were the heavy elements created?

\end{itemize}

There are several missions underway or planned, aimed at tackling some
of these questions. ESA's {\em Planck Surveyor Mission} (scheduled for
launch in 2007) will produce a detailed map of the spectrum of
fluctuations of the cosmic microwave background that will allow us to
tightly constrain some of the most fundamental cosmological parameters,
and will provide us with a better understanding of the physics that
governed the early Universe. Observation with {\em Herschel} (2007),
{\em NGST} (2009), and {\em ALMA} (2010) will provide us with
information about the formation of the first stars in the Universe, and
the formation and early evolution of galaxies.

However, in the course of the evolution of the Universe, most of the
baryonic matter must have been heated to temperatures in which it will
emit X-rays. Part of this matter, the hottest and denser part of it, is
readily detectable in clusters of galaxies, but a large fraction of it
at much lower densities, the warm-hot intergalactic medium
(\cite{jbleeker-D:cen99}), has not yet been observed. Massive accreting
black holes, which probably played a central role in the formation and
evolution of the galaxies that host them, are only observable in
X-rays. Hard X-rays can penetrate the thick clouds of gas and dust in
the centers of young galaxies, and therefore observations in these
wavelengths are needed to distinguish between energy output from star
formation and accretion in these objects.

All of these subjects require very sensitive X-ray instruments, with
sufficient angular resolution to avoid confusion at high redshifts, and
high energy resolution to be able to study in detail the physical
properties of these young objects.

\subsection{The first black holes}

Deep surveys with XMM-Newton and Chandra are beginning to show that the
fraction of galaxies harboring a massive ($\simmore 10^{6} \msun$)
black hole at its center is larger at redshifts larger than 1 to 3,
than it is in the local universe. Previous X-ray and infrared surveys
show that the star formation rate and the space density of AGN was a
factor of $\sim 100$ larger at redshifts of 2-5 than it is at present.
It is not yet clear what is the reason of this recent decline in the
star formation rate and the rate at which black holes at the centers of
galaxies accrete matter. However, these results indicate that the
universe was much more active at redshifts larger than 3 than it is
now, and that X-ray emission from those ages is of crucial importance
for the proper assessment of evolutionary scenarios.

The current paradigm of structure formation in the Universe
(\cite{jbleeker-D:peb74,jbleeker-D:whi78}) states that small-scale
objects forms first, and that they afterward merge to form larger ones.
Within this scenario, it is not clear whether black holes at the center
of active galaxies formed in situ, or whether they grew from accretion
of smaller black holes as the smaller galaxies merged to form larger
ones. In any case, the tight correlation between the mass of the
central black hole and the velocity dispersion of the stars in the
bulge of the host galaxy (\cite{jbleeker-D:fer00,jbleeker-D:geb00})
demonstrates a close relation between massive central black holes and
galaxy formation.

\begin{figure}[bt]
  \begin{center}
    \epsfig{file=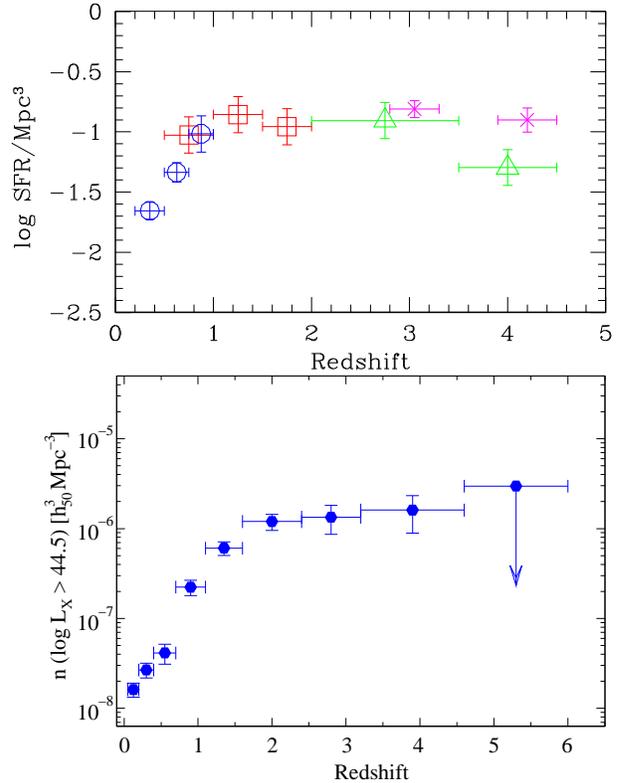, width=8cm, angle=0}
  \end{center}
\caption{Star formation rate based on UV/optical observations (upper
panel; from Steidel et al. (1999), and space density of X-ray selected
AGN with luminosities large than $3 \times 10^{44}$ erg s$^{-1}$ (lower
panel; from Miyaji et al. 2000), plotted as a function of redshift.}  
\label{jbleeker-D_fig:fig1}
\end{figure}

\begin{figure*}[bht]
  \begin{center}
   \epsfig{file=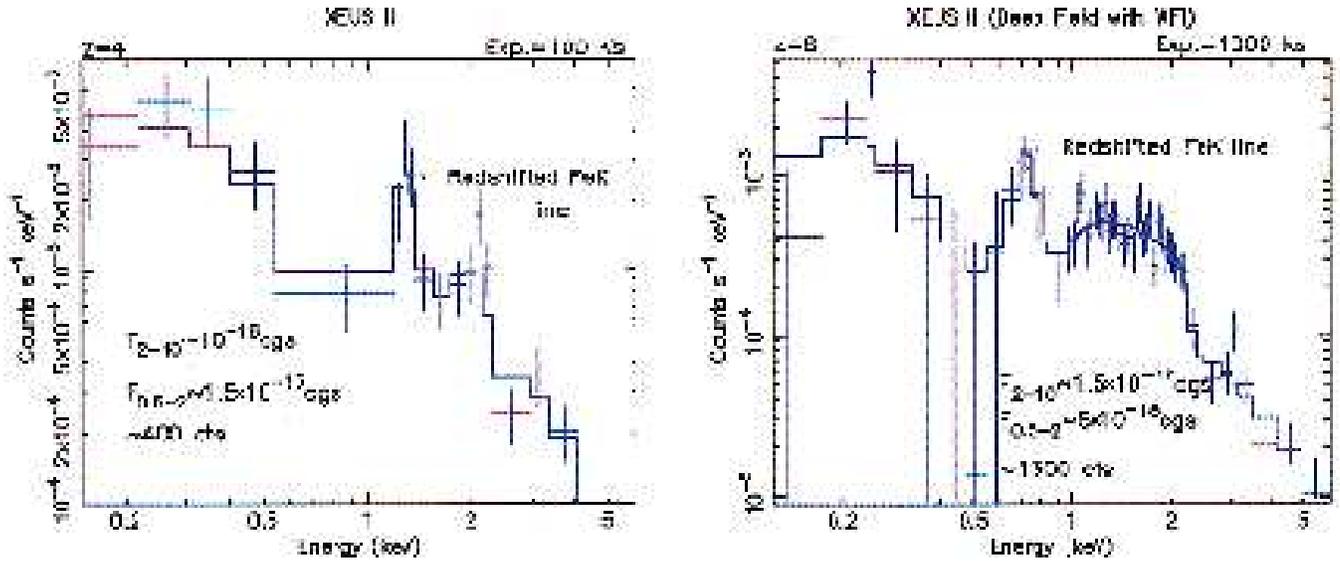, width=18cm, angle=0}
  \end{center}
\caption{XEUS 2 simulated spectra of a heavily absorbed starburst/AGN.
At low energies the input spectrum is dominated by thermal gas from a
starburst with $kT = 3$ keV, metallicity 30\,\% solar, and luminosity
$L_{\rm 0.5-2 keV} \sim 2 \times 10^{42}$ erg s$^{-1}$. At higher
energies the AGN dominates. The AGN luminosity is $L_{\rm 2-10 keV}
\sim 10^{44}$ erg s$^{-1}$; there is also a strong Fe line with
equivalent width $EW = 1$ keV. The intrinsic absorbing material has a
hydrogen column $N_{\rm H} = 10^{24}$ cm$^{-2}$. The left panel shows
the simulated spectrum for a source at redshift $z=4$, whereas the
right panel shows the same spectrum at a redshift $z=8$. Simulated
exposure times and observed fluxes are indicated in each panel
(reproduced from {\em The XEUS science case, ESA SP-1238}).}  
\label{jbleeker-D_fig:fig2}
\end{figure*}

Simple cosmological models of hierarchical formation predict the
existence of large population of quasars at higher redshifts. Although
original results seemed to indicate that star formation rate (and metal
production) peaked at a redshift around 1, new estimations that take
into account corrections for intrinsic absorption in quasars at high
redshifts indicate that the star formation rate increases up to $z \sim
1$, and remains more or less constant above this value
(\cite{jbleeker-D:ste99}). Evidence for a population of high-redshift,
highly absorbed, AGN is indicated by ROSAT results (Fig.
\ref{jbleeker-D_fig:fig1}), which show a more or less constant AGN
density from $z \sim 1$ to $z \sim 4$ (\cite{jbleeker-D:miy00}). This
is consistent with the fact that hard X-ray observations are much less
affected by absorption than optical and UV observations. This means
that a large fraction of the high-redshift Universe is obscured. 
Interalia, it should be noted that some models
(\cite{jbleeker-D:hai99}) and recent data (Hasinger, private
communication), seem to indicate that the space density of AGN might
decline again beyond $z \sim 4$. Typical estimates indicate that highly
obscured AGN will produce fluxes of $< 10^{-16}$ erg cm$^{-2}$ s$^{-1}$
in the 2--10 keV energy range. Neither Chandra nor XMM-Newton will be
able to measure X-ray spectra at such low flux levels. In order to
sample the high-redshift AGN a telescope with a larger spectroscopic
effective area is required.

If indeed black holes and AGN were formed at redshifts larger than 5, a
question remains as to whether the intergalactic medium at those
redshifts was ionized by AGN or by the first massive stars formed in
the early Universe, or whether the AGN could have influenced the
formation of structure.

Given its expected sensitivity (see below), XEUS will be capable of
addressing all these issues. XEUS will be able to detect AGN with
central black holes of $10^{6-7} \msun$ which emit at luminosities of
$10^{43-44}$ erg s$^{-1}$ at redshifts of 20, and to study their X-ray
spectra (via the detection of line emission from Fe K$_{\rm \alpha}$)
up to redshift of 10. Spectroscopic studies of the Fe line (Fig.
\ref{jbleeker-D_fig:fig2}) and variability studies of the X-ray
luminosity of the underlying galaxy will provide information about the
geometry of the accretion flow in the vicinity of a black hole, will
constrain the geometry close to the black hole's event horizon, and
will allow us to measure the mass, and possibly the spin, of the black
holes. By observing a large sample of AGN at different redshifts it
will be possible to study the evolution of black hole mass and spin
from the early Universe up till now. From the black-hole spin rate
history it will be possible to assess the way these black holes formed
and grew, given that steady accretion would yield high spin rates,
whereas black hole mergers would yield smaller spin rates.

\subsection{Groups and clusters of galaxies at high {\em z}}

In the standard cosmological scenario of hierarchical structure
formation, small-scale structures collapse first, and grow into larger
aggregates (\cite{jbleeker-D:whi78}). In a nutshell, dark matter
initially accretes into larger and larger halos due to the
gravitational amplification of the initial small density fluctuations.
Baryonic matter associated to these halos can dissipate energy and cool
down by radiation to form stars.

While observations of the CMB (e.g., with the Planck Surveyor Mission)
can provide the spectrum of the initial fluctuations, as well as the
global parameters governing cosmic evolution, they alone cannot provide
information about the physics of gas cooling and heating processes that
take place during galaxy formation.

As the baryonic matter collapses and cools, much of the gas will emit
soft X-rays. Therefore, it is only through X-ray observations that it
is possible to study the hot baryonic matter component within the large
scale structure of the Universe. It is through X-ray observations that
we have been able to trace the hot plasma bound to the gravitational
potentials of groups and clusters of galaxies, the largest mass
aggregates in the Universe. These studies have provided some of the
most fundamental constraints  to-date used to test cosmological models.
X-ray studies of clusters of galaxies have been used to measure the
ratio of dark matter to baryonic matter at different length scales,
which reflects the way large objects are formed through the
hierarchical merging of smaller units, but also has allowed us to trace
the abundance of iron and other heavy element in intergalactic space,
which seems to imply a much more violent star burst history than
previously assumed.

However, because of the limitations of XMM-Newton and Chandra, so far
these studies could only be carried out for objects in the local
Universe; similar studies at large redshifts, which could provide 
evidence for evolution from the early Universe up till now, require
much higher sensitivity than current missions can provide.  The large
effective area and high angular resolution of XEUS will enable these
studies to be extended to redshifts of $z \simmore 2$, to study groups
and clusters of galaxies at the epoch when these massive objects first
emerged (\cite{jbleeker-D:ven02}), presumably when the star formation
rate, and the production of heavier elements, peaked.

\begin{figure}[bt]
  \begin{center}
    \epsfig{file=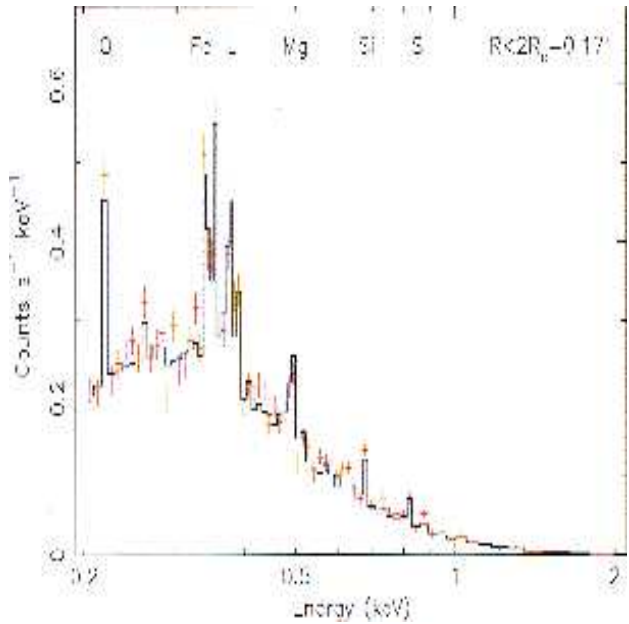, width=8.2cm, height=8.2cm, angle=0}
  \end{center}
\caption{Simulated spectrum of a small group with kT=1.5 keV and a
bolometric luminosity of $10^{43}$ erg s$^{-1}$ at a redhsift of z=2
(reproduced from {\em The XEUS science case}, ESA SP-1238).} 
\label{jbleeker-D_fig:fig3}
\end{figure}

These observations will provide a better understanding of the role of
feedback from the cluster galaxies to the physical and chemical state
of the gas at high $z$. For instance, the evolution of the
intra-cluster medium is not purely governed by gravitational effects.
Galaxies are injecting metals and energy, probably at early epochs via
supernova driven winds (\cite{jbleeker-D:mad01}); this feedback is
likely to affect cluster formation and evolution. Central cooling in
the early dense groups and cluster cores can have an important effect
on the evolution of these systems. Additionally, by tracing the outer
parts of clusters and studying the larger scale structures, such as
filaments, we expect to observe the growth of clusters of galaxies by
accretion of intergalactic matter.

Deep X-ray observations will test the hierarchical formation scenario
from merging activity at high $z$ and the evolution of sub-clustering
with $z$. For instance, the collision of two sub-clusters should be
manifest in temperature maps, through heated gas between the
sub-clusters before the collision as well as steep temperature
gradients (changes by factor of 2--3 over 200 kpc) at the shocks formed
during the collision. The velocities of the gas can range from
200--2000 km/s.

\begin{figure}[bt]
  \begin{center}
    \epsfig{file=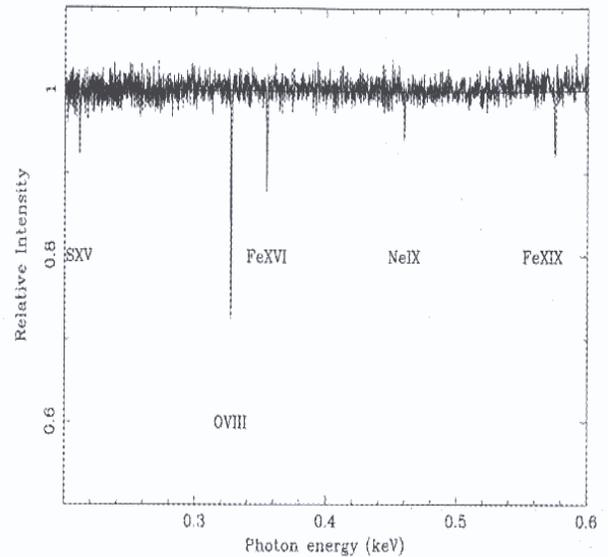, width=8.2cm, height=8.2cm, angle=0}
  \end{center}
\caption{Simulated XEUS-2 absorption-line spectrum of a small group of
galaxies in front of an AGN with an observed 0.5--2 keV flux of
$10^-12$ erg cm$^{-2}$ s$^{-1}$. The exposure time is 100 ksec. The
group is assumed to be at $z=1$, to have a core radius of 50 Kpc, and
0.1--2.4 keV luminosity of $10^{-42}$ erg s$^{-1}$ (reproduced from {\em
The XEUS science case}, ESA SP-1238).}  
\label{jbleeker-D_fig:fig4}
\end{figure}

It is important to notice that the history of the gas and galaxy
formation are deeply interconnected. During galaxy formation, the
temperature, density, and chemical composition of the gas are
fundamental in determining the fate of the collapsing gas: it is
critical whether the gas can cool or not. X-ray studies are the most
direct way to obtain information on the physical state of the gas,
which ultimately controls the overall history of galaxy formation. 

XEUS will have the angular resolution and sensitivity to study the
dynamics of the gas in those systems in detail, as well as the spectral
resolution to measure the mass motion from emission line spectroscopy.

But the hot plasma in clusters of galaxies is probably a small fraction
of all the baryonic matter in the Universe. Numerical simulations
predict (\cite{jbleeker-D:dav01}) that 30--40\,\% of the baryons are in
the warm-hot intergalactic medium (\cite{jbleeker-D:cen99}), and
therefore their radiation is too faint to be detected. However,
absorption features against a bright source in the background (e.g.,
gamma-ray bursts may in principle be detected at redshift of 10 or
more) can make them visible (Fig. \ref{jbleeker-D_fig:fig3}), as is
well known from the classical Lyman-$\alpha$ forest spectroscopy. If
the temperature of this matter is $> 10^{5}$ K, the corresponding
absorption features lie in the X-ray range, the ``X-ray forest''
(\cite{jbleeker-D:hel98,jbleeker-D:per98}).

Recent hydrodynamic cosmological simulations (\cite{jbleeker-D:hel98})
show that the strongest absorption features from the intergalactic
medium (IGM) should be produced by O-VII and O-VIII; for an IGM with a
metallicity 10\,\% solar, the absorbers that may be detectable given
Chandra and XMM-Newton sensitivities have temperatures in excess of $
10^{5.5} - 10^{6.5}$ K and overdensities $\delta \simmore 100$
(\cite{jbleeker-D:che02}). To be able to sample the IGM over a larger
range of temperatures and overdensities, and to be able to detect other
elements besides oxygen, a mission with the characteristics of XEUS is
needed (\cite{jbleeker-D:che02}).

\begin{figure}[bt]
  \begin{center}
    \epsfig{file=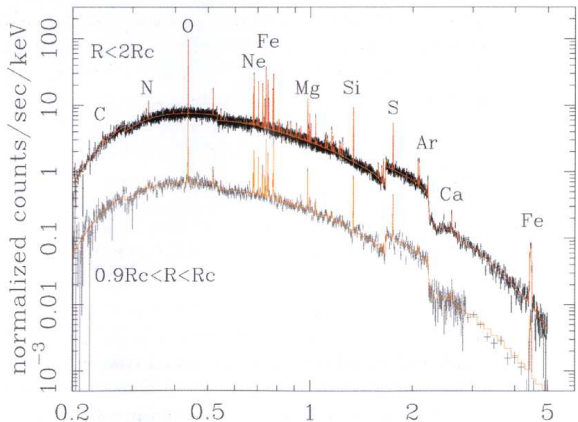, width=8.5cm, angle=0}
  \end{center}
\caption{Simulated XEUS-2 spectra of a cluster of galaxies at $z=0.5$
having a bolometric luminosity of $1.6 \times 10^{44}$ erg s$^{-1}$.
The upper and lower spectra represent simulated measurements from within
2 core radii, and an annulus between 0.9 and 1 core radius, respectively
(reproduced from {\em The XEUS science case}, ESA SP-1238).}
\label{jbleeker-D_fig:fig5}
\end{figure}

\subsection{Heavy elements enrichment history}

One of the fundamental issues in astrophysics today is how heavy
elements formed, and what is the evolution of the metallicity of the
intergalactic medium (\cite{jbleeker-D:fera00}). This issue is strongly
related to the star formation history, the possible variations of the
initial stellar mass function with environmental conditions, and the
circulation of matter between the different phases of the Universe.

This issue will be addressed in part by constraining the history of
star formation rate using NGST, Herschel, and ALMA. However, the X-rays
provide the best possible way of studying the history of heavy element
production. Clusters of galaxies are the largest closed systems where
the chemical enrichment process can be studied in detail. X-ray
observations of lines emitted by the hot intra-cluster medium can be
used to determine abundances up to much higher redshifts than possible
via optical observations of normal galaxies (Fig.
\ref{jbleeker-D_fig:fig4}), with the advantage that X-ray observations
provide direct measurement of the abundances, without having to rely on
the indirect indicators used in the optical to estimate the element
abundances in galaxies.

Abundance gradients in clusters and groups of galaxies can be used to
directly assess the chemical enrichment history of the intergalactic
medium, much better than would be possible using individual galaxies.
Abundances in very poor groups can be measured to any redshift using
X-ray absorption line spectroscopy as long as a bright background
source can be found. In the case of Gamma-ray burst afterglows this is
also true for the host galaxy and all intervening systems. XEUS will
measure the abundances of all astrophysically abundant elements, down
to the photon detection limit. Since XEUS' energy resolution will be
much better than the equivalent width of the strongest emission lines
over a large range of temperatures, it will be possible to obtain
information on the heavy element enrichment history of the intracluster
medium of a quality now only reached for our Galaxy. This result will
have direct implications on our understanding of the evolution of
cluster galaxies.

XEUS will make it possible to trace the evolution of the intra-cluster
medium abundances back to at least $z \sim 2$, and down to poor
clusters, therefore constraining the epoch of production and ejection
of the heavy elements, and unveiling the interplay between the
dynamical, in particular the effect of mergers, and chemical history of
clusters. By comparing the spatial distribution of heavy elements and
galaxies up to high redshifts, XEUS will provide a strong constraint on
ejection process, wind or ram pressure stripping. 

Furthermore, using XEUS it will be possible to probe abundances in the
intergalactic medium, and estimate the metal production efficiency of
field and cluster galaxies. It will also be possible to independently
constrain the star formation history, since the overall cluster metal
content is a fossilized integral record of the past star formation, and
to precisely measure the abundance relative to Fe back to $z \sim 2$, 
which can be used to assess the relative importance of type I and type
II supernovae and their past rates. This is a strong constraint on the
initial mass function, and also has far reaching consequences on our
understanding of the thermal history of the inter-cluster medium.

\begin{figure}[bt]
  \begin{center}
   \epsfig{file=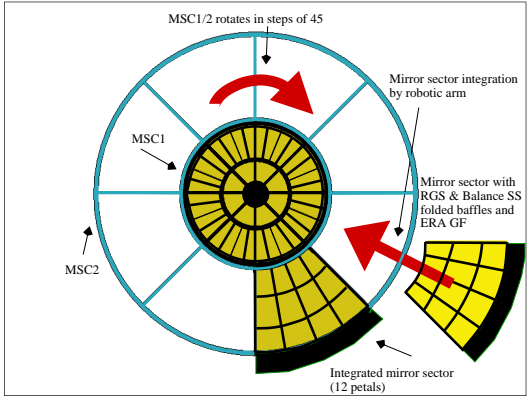, width=6cm, angle=-90}
  \end{center}
\caption{Schematic figure showing XEUS mirror growth (reproduced from
{\em The XEUS telescope}, ESA SP-1253).}  
\label{jbleeker-D_fig:fig6}
\end{figure}
\vspace{-0.2cm}

\section{The mission}

The key characteristics of XEUS are large X-ray mirror aperture, high
spectral resolution over wide-band energy range, and good angular
resolution (see Table \ref{jbleeker-D_tab:tab1}). With an area larger
than 20m$^{2}$ and an energy resolution of 2 eV below 2 keV, the XEUS
final configuration (see below) will be able to significantly detect
the most prominent X-ray emission lines of O-VII, SI-XIII and Fe-XXV
against the sky background and source continuum, while an angular
resolution of 2--5 arcsec will help minimize source confusion and will
reduce the background due to galactic foreground X-ray emission.

XEUS will consist of two separate spacecrafts, the mirror and the
detector spacecraft, respectively, injected in low-Earth orbit with an
inclination equal to that of the ISS (fellow-traveler orbit). The
mirror and detector spacecrafts will fly in formation, yielding a 50-m
focal length. The detector spacecraft will track the focus of the X-ray
telescopes with a precision of $\pm 1$ mm per degree of freedom.
Because the detector spacecraft will be flying in a non-Keplerian
orbit, it will need active orbit control.

XEUS is a two-step mission that will grow in space. XEUS-1, which will
be launched by an Ariane 5 rocket, will have a primary mirror of 4.5-m
in diameter comprising two concentric rings with so-called petal
structure, each petal consisting of a set of heavily stacked thin
mirror plates with a Wolter I type geometry. After five years in space,
the mirror spacecraft will dock with the ISS, where the European
Robotic Arm will be used to add additional mirror segments around the
central core, so the XEUS mirror will grow to its final size, 10-m in
diameter (Fig. \ref{jbleeker-D_fig:fig6}). The construction of the
mirrors poses several technological challenges, among them the
manufacturing of the mirror plates and their integration into modules,
and the integration of the individual modules in space into the final
configuration ensuring the required angular resolution of 2 to 5
arcsec. Alternative mission profiles, e.g. Ariane 5 launch of the
complete XEUS in low-Earth orbit, are also being studied in case the
necessary ISS infrastructure is not available 

The instrument payload model presently comprises a Wide-Field Imager
(WFI), and two Narrow-Field Cameras (NFC), one of them optimized for
soft X-rays and the other for hard X-rays. The WFI is based on pn-CCD
technology, with an energy resolution of 50 eV at 1 keV. The WFI CCD
covers $5\arcmin \times 5\arcmin$, and has a count rate capability with
low pile-up ($\simless 5\,\%$ of up to $\sim 1000$ count s$^{-1}$
within the PSF (HEW). The NFC is based on Superconducting Tunnel
Junctions (the soft-X camera) and Transition Edge Sensors (the hard-X
camera) technologies, will cover $30\arcsec \times 30\arcsec$, and will
allow for a time resolution of less than 5 microseconds. The energy
resolution will be better than 2 eV at 1 keV for the soft NFC, and 2 eV
and 5 eV, at 1 keV and 8 keV, respectively for the hard NFC. The high
energy resolution required seems feasible, given that at present a
single pixel micro-calorimeter can achieve an energy resolution of 3.9
eV at 5.9 keV, with a thermal response of 100 microseconds.

\begin{table}[bt]
\caption{XEUS 1/2 design specifications}
\begin{center}
\begin{tabular}{lcc}
\hline
                              & XEUS-1         & XEUS-2     \\
\hline
\hline
Energy range                  & 0.05 -- 30 keV &            \\
\hline
Effective area @ 1 keV        & 6 m$^{2}$      & 30 m$^{2}$ \\
\hline
Effective area @ 8 keV        & 3 m$^{2}$      &            \\
\hline
Angular resolution (HEW)      & 5''            &  2''       \\
\hline
\hline
\hline
\end{tabular}
\end{center}
Specifications that do not change in XEUS-2 are not indicated.
\label{jbleeker-D_tab:tab1}
\end{table}

When XEUS grows to its final configuration, the detector spacecraft
will be replaced with a new generation of instrument technology. XEUS
expected lifetime is 25 years or more.

The main characteristics of XEUS are indicated in Table 1. Given those
specifications, XEUS will be able to detect a source at a flux of $4
\times 10^{-18}$ erg cm$^{-2}$ s$^{-1}$ in the 0.5--2 keV energy range
in a 100 ks exposure, and it will be capable of measuring spectra down
to a flux of $10^{-17}$ erg cm$^{-2}$ s$^{-1}$. In order to achieve
these fluxes, the Half-Energy Width of the PSF must be $\simless 2-5$
arcsec.

\section{Present status and technology developments}

Several studies relating to XEUS enabling technologies are currently in
progress or have been planned for the near future. An 18-month system
level study will be carried out by the ESA directorate for Science and
the directorate for Manned Space and Microgravity, with technical
support from ISAS. This activity will kick off in April 2002 and will
address several basic feasibility issues concerning telescope
configuration, orbit, ISS interfacing and flight implementation
scenarios. More specific studies, targeted at formation flying (station
keeping), robotic assembly techniques, optical/IR straylight
suppression filters, and 50-milli-Kelvin closed cycle coolers are now
being addressed in the ESA Technology Research Programme (TRP). Under
the TRP, two X-ray sensor study contracts, one related to the
Wide-Field active pixel detector array and one dealing with a
squid-read-out X-ray bolometer/TES array, will be started at the
beginning of 2002, based on quite promising single-pixel results
obtained in various European (space) research institutes.

The major development item for the coming years is obviously the
technology, metrology, assembly, and active alignment of the X-ray
mirror petal. Starting from the XMM replica technology, several
alternative routes for manufacturing are now being pursued, including
the application of new, lightweight, materials like slumped glass and
SiC. Since the mirror technology is at the very heart of the XEUS
feasibility, it has been designated as a core technology development in
the ESA science programme, implying  major development funding in the
period 2002-2004. These efforts will be part of the so-called Core
Technology Programme, which was recently established within ESA science
for strategic technology investments.

\begin{acknowledgements}

We like to acknowledge the members of the XEUS Steering committee and
XEUS Astrophysics working group who have put together the science and
mission case as described in this overview paper.

\end{acknowledgements}

\end{document}